\documentclass[pra, aps,twocolumn,nofootinbib, groupedaddress, superscriptaddress,10pt, preprintnumbers]{revtex4-1}
\usepackage[dvipsnames]{xcolor}
\usepackage{amsmath}
\usepackage{amsfonts}
\usepackage{amssymb} 
\usepackage[hidelinks]{hyperref}
\usepackage[normalem]{ulem}
\usepackage{cleveref}

\def\fkg{\mathfrak{g}}
\def\bbZ{\mathbb{Z}}

\makeatother
\hypersetup{
    pdftitle={String Universality and Non-Simply-Connected Gauge Groups in 8d},
    pdfauthor={Mirjam Cvetic, Markus Dierigl, Ling Lin, Hao Y. Zhang},
    pdfsubject={Swampland, String Universality, Higher-form symmetries, Supergravity, F-theory}
}

\begin{document}

\preprint{\texttt{CERN-TH-2020-138, UPR-1306-T}}

\title{
	String Universality and Non-Simply-Connected Gauge Groups in 8d
	}

\author{Mirjam Cveti\v{c}} \affiliation{Department of Physics and Astronomy, University of Pennsylvania, Philadelphia, PA 19104-6396, USA}
  \affiliation{Center for Applied Mathematics and Theoretical Physics, University of Maribor, Maribor, Slovenia}
\author{Markus Dierigl} \affiliation{Department of Physics and Astronomy, University of Pennsylvania, Philadelphia, PA 19104-6396, USA} 
\author{Ling Lin} \affiliation{Department of Theoretical Physics, CERN, 1211 Geneva 23, Switzerland}
\author{Hao Y.~Zhang} \affiliation{Department of Physics and Astronomy, University of Pennsylvania, Philadelphia, PA 19104-6396, USA}

\begin{abstract}
\noindent 
We present a consistency condition for 8d ${\cal N} = 1$ supergravity theories with non-trivial global structure $G/Z$ for the non-Abelian gauge group, based on an anomaly involving the $Z$ 1-form center symmetry. The interplay with other Swampland criteria identifies the majority of 8d theories with gauge group $G/Z$, which have no string theory realization, as inconsistent quantum theories when coupled to gravity. While this condition is equivalent to geometric properties of elliptic K3 surfaces in F-theory compactifications, it constrains the unexplored landscape of gauge groups in other 8d string models.
\end{abstract}

\maketitle

\section{Introduction}

One of the important lessons from string theory is that consistency conditions of quantum gravity are highly restrictive.
In the low-energy limit, they result in a small and possibly finite subset of effective descriptions, leaving behind a vast ``Swampland'' of seemingly consistent quantum field theories coupled to gravity \cite{Vafa:2005ui}.
Recent attempts to specify the Swampland's boundary (cf.~\cite{Brennan:2017rbf,*Palti:2019pca} for reviews) have reinforced the idea of String Universality: every consistent quantum gravity theory is in the string landscape.

Prototypical examples of String Universality appear in eleven and ten dimensions, where low energy limits of M- and string theory give rise to the only consistent supergravity theories.
In ten dimensions (10d), this requires more subtle field theoretic arguments \cite{Adams:2010zy}, or the incorporation of extended dynamical objects in the theory \cite{Kim:2019vuc}, to ``drain'' the 10d supergravity Swampland.

In lower dimensions, one observes a broader spectrum of string-derived supergravity theories, but these nevertheless show some intricate structures not naively expected from field theory considerations.
For example, the rank $r_G$ of the gauge group in known string compactifications is bounded by $r_G \leq 26 - d$ in $d$ dimensions, and satisfies $r_G \equiv 1 \mod 8$ and $r_G \equiv 2 \mod 8$ in $d=9$ and $d=8$, respectively.
Likewise, not all gauge algebras have string realizations.
In particular, there are no string compactifications to 8d with $\mathfrak{so}(2n+1) \, (n\geq 3)$, $\mathfrak{f}_4$ and $\mathfrak{g}_2$.
Again, novel Swampland constraints \cite{Kim:2019ths,Montero:2020icj} and refined anomaly arguments \cite{Garcia-Etxebarria:2017crf} reproduce these restrictions, thus downsizing the 9d and 8d Swampland considerably.\footnote{As $\fkg_2$ does not suffer similar anomalies, it remains an open question if it truly belongs to the 8d Swampland.}

The goal of this work is to provide similar constraints for the \emph{global structure} of the gauge group of 8d ${\cal N}=1$ theories, by deriving a field theoretic consistency condition for the gauge group to take the form $G/Z$, with $Z \subset Z(G)$ a discrete subgroup of the center of $G$.
Taking inspiration from F-theory \cite{Vafa:1996xn,*Morrison:1996na,*Morrison:1996pp}, where the gauge group structure is encoded in the Mordell--Weil group of the elliptically-fibered compactification space \cite{Aspinwall:1998xj,Mayrhofer:2014opa,Cvetic:2017epq}, it appears that the allowed gauge groups $G/Z$ are heavily restricted. 
For example, there are no 8d string compactifications, including constructions beyond F-theory, that have gauge group $SU(n)/\bbZ_n$, whereas $SU(n)$ groups are ubiquitous.

These restrictions are mathematically well known from the classification of elliptic K3 surfaces \cite{MirandaPersson,*MirandaPersson_extremal,Shimada} (see also \cite{Hajouji:2019vxs}).
Focusing on $G$ a simply-connected non-Abelian Lie group\footnote{
More precisely, the most general gauge group is $\frac{G \times U(1)^r}{Z \times Z_\text{f}}$, with $Z \subset Z(G)$, i.e., $Z \cap U(1)^r = \{ 1 \}$.
In this work we consider constraints for $Z$ exclusively, leaving a more detailed study including $Z_\text{f} \subset Z(G \times U(1)^r) \cong Z(G) \times U(1)^r$, based on \cite{Cvetic:2017epq}, for future work. \label{footnote:1}
}, the geometry restricts $Z$; e.g., when $Z \cong \bbZ_\ell$, then $\ell\leq 8$.
Moreover, for each of the cases $\ell=7,8$, there is exactly one elliptic K3 on which F-theory compactifies to an 8d theory with $G = SU(7)^3/\bbZ_7$ and $[SU(8)^2 \times SU(4) \times SU(2)]/\bbZ_8$, respectively.
Analogous restrictions on gauge groups also appear in heterotic compactifications \cite{Font:2020rsk}.

A natural question is, whether these restrictions reflect limitations of string theory, or previously unknown consistency conditions of quantum gravity in 8d.

In this work, we show that the latter is the case.
The key is to realize a non-simply-connected group $G/Z$ by gauging the $Z$ 1-form center symmetry \cite{Kapustin:2014gua, Gaiotto:2014kfa}.
Thus, charting the Swampland of gauge groups $G/Z$ (in any dimension) can be equivalently tackled by studying consistency conditions for gauging $Z$ in gravitational theories.
As we will discuss below, in 8d ${\cal N}=1$ theories, one such condition is the absence of a mixed anomaly between the center 1-form symmetries and gauge transformations of higher-form supergravity fields, which would obstruct the gauging of $Z$.
This rules out a vast set of seemingly acceptable 8d ${\cal N}=1$ theories without known string constructions, and, in particular, reproduce the geometric restrictions in models with F-theory realization, thus providing further evidence for String Universality in 8d.

The anomaly originates from a generalization of the familiar $\theta$-term, $\theta \, \text{Tr}(F^2)$, in 4d.
There, the fractional shift of the instanton density $\text{Tr}(F^2)$, due to the presence of a background field for the $Z$ 1-form symmetry, breaks the $2\pi$-periodicity of $\theta$ \cite{Kapustin:2014gua,Gaiotto:2014kfa,*Gaiotto:2017yup,*Gaiotto:2017tne,*Cordova:2019jnf,Cordova:2019uob,Cordova:2019bsd}.
In higher dimensions, $\text{Tr}(F^2)$ couples to higher-form fields (e.g., to vector fields in 5d and tensors in 6d), which themselves possess gauge symmetries.
These can lead to mixed anomalies with the $Z$ 1-form center symmetry \cite{Apruzzi:2020zot, BenettiGenolini:2020doj}.\footnote{See also \cite{Morrison:2020ool, *Albertini:2020mdx,*Closset:2020scj,*DelZotto:2020esg,*Bhardwaj:2020phs} for recent treatments of higher-form symmetries in higher-dimensional setups and \cite{Dierigl:2020myk} for an analysis of the global gauge group in 6d SCFTs.}

The analogous coupling in 8d involves a 4-form $B_4$.
Crucially, while such a term is absent in a pure 8d supersymmetric gauge theory (as there are no appropriate fields $B_4$ in the ${\cal N}=1$ vector multiplet), the coupling $\sum_i m_i B_4 \wedge \text{Tr}(F_i^2)$ necessarily exists if one includes a gravity multiplet, which contains a unique tensor $B_2$ that is dual to $B_4$ \cite{Awada:1985ag}.
Supersymmetry further demands that $m_i \neq 0$, \cite{Salam:1985ns}.
A mixed anomaly involving the symmetries of $B_4$, which must be gauged, and the center 1-form symmetry $Z$ can, therefore, obstruct the gauging of the latter.
The vanishing of this anomaly is, hence, a \emph{necessary} condition to obtain a non-simply-connected gauge group $G/Z$.
Remarkably, in models with $m_i=1$ this condition turns out to reproduce geometric properties of elliptic K3 manifolds!
In combination with other Swampland criteria that constrain the coefficients $m_i$, this anomaly restricts possible combinations of simply-connected $G = \prod_i G_i$ and $Z \subset Z(G)$ in 8d. 
With this, we can consequently ``drain'' large portions of the 8d Swampland, and make predictions in corners of theory space where the global gauge group structure in corresponding string models is yet to be explored.

\section{Mixed anomaly for center symmetries in 8d supergravity}
\label{sec:anomaly}

Let $G = \prod_i G_i$ be a non-Abelian group, where $G_i$ are simple simply-connected Lie groups with algebra $\fkg_i$.
In 8d ${\cal N}=1$, the gauge potential $A_i$, with field strength $F_i$, of the $\fkg_i$ gauge symmetry comes in a vector multiplet with adjoint fermions.
There are no other massless charged matter states, so at low energies one expects a discrete $Z(G) = \prod_i Z(G_i)$ 1-form symmetry \cite{Gaiotto:2014kfa}.
Moreover, since the only massless fermions transform in a real representation, there are no pure gauge anomalies \cite{Taylor:2011wt}.

Besides the vector multiplets, 8d ${\cal N}=1$ supergravity contains the gravity multiplet with a 2-form gauge field $B_2$ as one of its component fields \cite{Salam:1985ns}. 
The field strength $H_3$ of this 2-form field obeys a modified Bianchi identity involving the gauge fields of the theory,
\begin{align}
H_3 = d B_2 + \sum_{i} m_i \, \text{CS}(A_i) \,.
\end{align}
Here, $\text{CS}(A_i)$ are the Chern--Simons functionals for the gauge factor $G_i$.

The positive integers $m_i$ associated with each gauge factor $G_i$, which we will refer to as the ``level'' of $G_i$, are a priori free parameters of the supergravity theory.
They can be interpreted as the magnetic charge of gauge instantons under $B_2$ --- more apparent in the dual formulation, with $B_2$ replaced by its magnetic-dual 4-form $B_4$. The most general Lagrangian contains the coupling \cite{Awada:1985ag}
\begin{align}
	\int_{M_8} \sum_i B_4 \wedge m_i \, \text{Tr}(F_i \wedge F_i) =: \int_{M_8} \sum_i B_4 \wedge m_i \, I_4(G_i)\, ,
	\label{eq:anomcoup}
\end{align}
where the trace is normalized such that the instanton density $I_4(G_i) = 1$ for a one-instanton configuration of a $G_i$-bundle on a 4-manifold $M_4$. 

The center 1-form symmetry of $G_i$ can be coupled to a 2-form background field $C_2^{(i)}$ which takes values in $Z(G_i)$.
When $C_2^{(i)}$ is non-trivial, it twists the $G_i$-bundle into a $G_i/Z(G_i)$-bundle with second Stiefel--Whitney class $w_2(G_i/Z(G_i)) = C_2^{(i)}$ \cite{Kapustin:2014gua,Gaiotto:2014kfa} that contributes to \eqref{eq:anomcoup},
\begin{align}
I_4(G_i/Z(G_i)) \equiv \alpha_{G_i} \mathfrak{P}\big(C^{(i)}_2\big) \mod \bbZ \, ,
\end{align}
with $\mathfrak{P}$ the Pontryagin square.
This contribution is, in general, fractional due to the coefficients $\alpha_{G_i}$ derived in \cite{Cordova:2019uob}, which we reproduce here:
\begin{center}
\renewcommand*{\arraystretch}{1.4}
\begin{tabular}{| c | c | c |}
\hline
$G_i$ & $Z(G_i)$ & $\alpha_{G_i}$ \\ \hline \hline
$SU(n)$ & $\mathbb{Z}_n$ & $\tfrac{n-1}{2n}$ \\ \hline
$Sp(n)$ & $\mathbb{Z}_2$ & $\tfrac{n}{4}$ \\ \hline
$Spin(2n+1)$ & $\mathbb{Z}_2$ & $\tfrac{1}{2}$ \\ \hline
$Spin(4n+2)$ & $\mathbb{Z}_4$ & $\tfrac{2n+1}{8}$ \\ \hline
$Spin(4n)$ & $\mathbb{Z}_2^{(L)} \times \mathbb{Z}_2^{(R)}$ & $\big( \tfrac{n}{4}, \tfrac{1}{2}\big)$ \\ \hline
$E_6$ & $\mathbb{Z}_3$ & $\tfrac{2}{3}$ \\ \hline
$E_7$ & $\mathbb{Z}_2$ & $\tfrac{3}{4}$ \\ \hline
\end{tabular}
\end{center}

Analogous to the situation in 6d \cite{Apruzzi:2020zot}, the coupling \eqref{eq:anomcoup} combines the fractional instanton configuration with a large gauge transformation $B_4 \rightarrow B_4 + b_4$, with $b_4$ a closed 4-form with integer periods, into a phase $2\pi i {\cal A}(b_4, C^{(i)}_2)$ for the partition function
\begin{align}\label{eq:anomaly}
	{\cal A}(b_4, C^{(i)}_2) = \sum_i m_i \alpha_{G_i} \int_{M_8} b_4 \cup \mathfrak{P}(C^{(i)}_2) \, .
\end{align}
While $\int_{M_8} b_4 \cup \mathfrak{P}(C^{(i)}_2) \in \bbZ$ for arbitrary $b_4$, the whole expression is, in general, fractional due to $\alpha_{G_i}$.
By generalizing the arguments presented in \cite{Hsieh:2020jpj,Apruzzi:2020zot}, the electrically charged objects for $B_4$ would acquire a fractional charge if this anomalous phase is non-trivial.
Since this violates charge quantization, the fractional shift \eqref{eq:anomaly} cannot be compensated and can be understood as an anomaly between the large gauge transformations of $B_4$ and the center 1-form symmetries.
As the former symmetry is gauged, one cannot allow for background fields $C_2^{(i)}$ where \eqref{eq:anomaly} is non-trivial.
Similar to the 6d setting \cite{Apruzzi:2020zot}, we expect that the violation of charge quantization is tied to the lack of counterterms that could absorb this anomaly. 
Moreover, we expect that arguments developed in \cite{Cordova:2019bsd} suggest that there cannot be a topological Green--Schwarz mechanism that cancels the above anomaly.\footnote{
Note that \cite{Cordova:2019bsd} discusses precisely the 4d analog of the anomaly \eqref{eq:anomaly} involving the $\theta$-angle instead of $B_4$.}

In general, while the individual centers $Z(G_i)$ are anomalous, there can be a non-trivial subgroup $Z \subset \prod_i Z(G_i)$ that is anomaly-free.
Assuming that there are no other obstructions to switch on a background for this subgroup $Z$ of the center, or other breaking mechanisms, this combination should be gauged, in line with common lore that in quantum theories of gravity no global symmetries (including discrete and higher-form symmetries) are allowed \cite{Brennan:2017rbf,*Palti:2019pca,Harlow:2018tng}.
In turn, this leads to the gauge group $G/Z$.

\subsection{Condition for Anomaly-Free Center Symmetries}

In the following, we will discuss how to determine subgroups $\bbZ_\ell \cong Z \subset Z(G)$, for which a 1-form symmetry background has no fractional contribution \eqref{eq:anomaly} --- a necessary condition to gauge $Z$.

Let $Z(G) = \prod_{i=1}^s \bbZ_{n_i}$, and $(k_1,...,k_s) \in \prod_{i=1}^s \bbZ_{n_i}$ be the generator for $Z \cong \bbZ_\ell$.
This means that $\ell$ is the smallest integer such that $k_i \ell \equiv 0 \mod n_i$ for all $i$.
The generic background for the $Z(G)$ 1-form symmetry consists of fields $C_2^{(i)}$ for each $\bbZ_{n_i}$ factor of $Z(G)$.
Specifying a specific background for a subgroup then amounts to correlating the a priori independent $C_2^{(i)}$'s \cite{Cordova:2019uob}.
In particular, the background $C_2$ for $Z \cong \bbZ_\ell$ corresponds to setting $C_2^{(i)} = k_i C_2$.

For concreteness, let $G = \prod_{i=1}^s SU(n_i)$.
Then, the anomalous phase \eqref{eq:anomaly} in a non-trivial $C_2$ background of the subgroup $Z \subset Z(G)$ is
\begin{align}
\mathcal{A}(b_4, C_{2}^{(i)}) 
	= \,  \left( \sum_{i=1}^s \frac{n_i - 1}{2n_i} k_i^2 m_i \right) \int_{M_8} b_4 \cup \mathfrak{P}(C_2) \, ,
\end{align}
where we used $\mathfrak{P}(k C) = k^2 \mathfrak{P}(C)$.
Thus, the anomaly vanishes if the coefficient is integral.

Note that the anomaly contribution of non-$SU$ groups can be written as a sum of contributions from $SU(n)$-subgroups \cite{Cordova:2019uob}.
Therefore, by further restricting ourselves to rank$(G)\leq 18$ (which is the 8d bound for the total gauge rank \cite{Montero:2020icj}), we can exhaustively scan for all possible groups $G$ that have an anomaly-free $\bbZ_\ell \subset Z(G)$ with given $\ell$, by finding $s$ triples of integers $(n_i, k_i, m_i)$ such that
\begin{align}\label{eq:anomaly_product_of_su}
	\sum_{i=1}^s \frac{n_i - 1}{2n_i} \, k_i^2 m_i \in \bbZ \, , \quad \text{with} \quad k_i \cdot \ell \equiv 0 \mod n_i \, .
\end{align}

Clearly, the levels $m_i$ play an important role.
From an effective field theory perspective, these are free parameters that define the theory. However, these parameters themselves are constrained by Swampland criteria.
By the Completeness Hypothesis \cite{Polchinski:2003bq,*Banks:2010zn}, the 2-form field $B_2$ couples to strings which carry localized degrees of freedom sensitive to the gauge group. 
These left-moving, charged excitations on the string have to cancel the worldvolume anomalies arising due to anomaly inflow \cite{Kim:2019vuc, Katz:2020ewz}. However, in $d$ dimension the left-moving central charge for such a string is bounded by $c_L \leq 26 - d$. 
While each $U(1)$ gauge factor contributes to $c_L$ with $c_{U(1)}=1$, each non-Abelian simple gauge factor $G_i$ with level $m_i$ contributes $c_i = \tfrac{m_i \, \text{dim}(G_i)}{m_i + h_i}$, with $h_i$ the dual Coxeter number of $G_i$.
Hence, we have
\begin{align}
\sum_{i} \frac{m_i \, \text{dim}(G_i)}{m_i + h_i} + n_{U(1)} \leq 18 \, .
\label{eq:anominfl}
\end{align}
Combined with the constraint that the rank of the total gauge group of the 8d supergravity theory can be only 2, 10, or 18 \cite{Montero:2020icj}, the $m_i$ are considerably restricted.
In particular, it is easily shown that in the rank-18 case, all $m_i$ must be 1 and all non-Abelian factors must have simply-laced algebras (see appendix \ref{app:m} for more details).
This is well-known in string compactifications, where $m_i$ are the levels of the worldsheet current algebra realizations of spacetime gauge groups, and are all $m_i=1$ on the rank-18 branch of the ${\cal N}=1$ moduli space.
As we will see now, the anomaly matches known geometric limitations in the F-theory realization of 8d rank-18 theories, which restricts the possible global gauge group structures. 
In the lower-rank cases these conditions can constrain gauge groups, whose algebras and levels fit in constructions such as the CHL string \cite{Chaudhuri:1995fk}, but whose global structure is yet to be explored.

\subsection{Anomaly-Free Centers in Theories of Rank 18}
\label{sec:singlefac}

All rank-18 ${\cal N}=1$ supergravity theories with a known string origin have a construction via F-theory \cite{Vafa:1996xn}, where physical features, including the global gauge group structure, are encoded in the geometry of elliptically-fibered K3 surfaces \cite{Taylor:2011wt,Weigand:2018rez,Cvetic:2018bni}.
In particular, there are beautiful arithmetic results \cite{MirandaPersson,*MirandaPersson_extremal} which asserts that F-theory compactifications with non-Abelian gauge group $G/Z$, where $G$ consists only of $SU(n_i)$ factors, must satisfy 
\begin{align}\label{eq:anomaly_cancellation_rank18}
	\sum_{i=1}^s \frac{n_i -1}{2n_i} k_i^2 \equiv 0 \mod \bbZ \,.
\end{align}
with $(k_1,...,k_s) \in \prod_i Z(SU(n_i))$ the generator of any $\bbZ_\ell \subset Z$ subgroup.

While we defer a more detailed explanation of the geometric origin to this formula to appendix \ref{sec:geometry}, it is obvious that it fully agrees with the cancellation condition for every $\bbZ_\ell$ subgroup of the center 1-form symmetry \eqref{eq:anomaly_product_of_su}, as for rank-18 theories all levels are fixed to $m_i = 1$. 
We therefore find a deep connection between the mixed anomaly of the supergravity theory and the geometrical properties of F-theory compactifications.

The constraint is particularly powerful when the order $\ell$ of the gauged center subgroup is the power of a prime number.
For such $\ell \geq 9$, one can show that there are no possible sets $\{(n_i,k_i)\}$ for which the anomaly vanishes with gauge groups of rank $\leq 18$.
For $\ell=7$, there is exactly one configuration with three simple non-Abelian factors, $n_1 = n_2 = n_3 = 7$ and $(k_1,k_2,k_3) = (1,2,3)$, corresponding to an $SU(7)^3/ \bbZ_7$ theory.
This agrees with the classifications of K3 surfaces \cite{MirandaPersson_extremal} for F-theory constructions as well as possible heterotic realizations \cite{Font:2020rsk}.
Likewise, in the case $\ell = 8 =2^3$, the 1-form anomaly \eqref{eq:anomaly_cancellation_rank18} allows for only $G = SU(8)_1 \times SU(8)_2 \times SU(4) \times SU(2)$, into which the $\bbZ_8$ sub-center embeds as $(k_1, k_2, k_{SU(4)}, k_{SU(2)}) =(1,3,1,1)$.
Furthermore, if we also take inspiration from geometric properties of K3 surfaces --- there always is one $SU(n_i)$ factor with $\ell$ dividing $n_j$ --- we can show that there are \emph{no} possible configurations $(n_i,k_i)$ for all $\ell \geq 10$.
This also matches the dual heterotic constructions \cite{Font:2020rsk}.

\subsection{Predictions for Simple Groups}

To further showcase the constraining power of the field-theoretic anomaly argument, we use \eqref{eq:anomaly} to rule out 8d ${\cal N}=1$ theories with gauge group $G/Z$, where $G$ is a simple Lie group and $Z\subset Z(G)$ a non-trivial subgroup.
For $G$ with $m=1$ and rank$(G) \leq 18$, any $G/Z$ is inconsistent except 
\begin{align}
		& \frac{SU(16)}{\bbZ_2} \, , \quad \frac{SU(18)}{\bbZ_3} \, , \quad  \frac{Spin(32)}{\bbZ_2} \, , \label{eq:allowed_groups1}  \\
		& \frac{Sp(4)}{\bbZ_2} \, , \quad \frac{Sp(8)}{\bbZ_2} \, , \quad \frac{SU(8)}{\bbZ_2} \, , \quad \frac{SU(9)}{\bbZ_3} \, , \label{eq:allowed_groups2}  \\
		& \frac{Spin(16)}{\bbZ_2} \, , \quad \frac{Sp(12)}{\bbZ_2} \, , \quad \frac{Sp(16)}{\bbZ_2} \, . \label{eq:allowed_groups3}
\end{align}
The groups \eqref{eq:allowed_groups1} indeed correspond to the only cases with simple $G$ realizable via F-theory on elliptic K3's.
The groups in \eqref{eq:allowed_groups2} are subgroups of $Sp(10)$, which at $m=1$ can be constructed from the CHL string \cite{Chaudhuri:1995fk}.
Note that this rules out all other $Sp(k)/\bbZ_2 (k<10)$ theories, which seemed consistent based on the perturbative CHL spectrum \cite{Mikhailov:1998si}.
As we are not aware of any systematic study of the global gauge group structure in CHL compactifications, we view this as a prediction based on the 1-form anomaly \eqref{eq:anomaly}, which is also consistent with other Swampland arguments \cite{Montero:2020icj}.
Groups in \eqref{eq:allowed_groups3} have no known 8d string realization at $m=1$.
However, while $Sp(12)/\bbZ_2$ and $Sp(16)/\bbZ_2$ are excluded at any $m$ due to the bound \eqref{eq:anominfl}\footnote{In particular, \eqref{eq:anominfl} provides a physical explanation to the limitation $k \leq 10$ for $\mathfrak{sp}(k)$ gauge algebras known in 8d string constructions.}, $Spin(16)/\bbZ_2$ does arise at $m=2$ as a Wilson line reduction of the $E_8$ CHL string.

More generally, at level $m=2$, the center anomaly in conjunction with the bound \eqref{eq:anominfl} can rule out all $G/Z$ theories with simple $G$ except for
\begin{align}
	\begin{split}
		& \frac{SU(4)}{\bbZ_2} \, , \quad \frac{SU(8)}{\bbZ_2} \, , \quad \frac{SU(9)}{\bbZ_3} \, ,\quad \frac{Sp(2)}{\bbZ_2} \, , \quad \frac{Sp(4)}{\bbZ_2} \, ,  \\
		& \frac{Spin(8)}{\bbZ_2} \, , \quad \frac{Spin(16)}{\bbZ_2} \, , \quad SO(2n) \quad \text{with} \quad 2 \leq n \leq 9 \, ,
	\end{split}
\end{align}
all of which could, in principle, arise in CHL compactifications \cite{Mikhailov:1998si}.
We will leave an explicit verification and analysis of the global gauge group in these types of 8d string models for future works.
Note that $SO(2n)$ ($n$ odd) and $Sp(2)/\bbZ_2$ seem to be ruled out in 8d by independent Swampland arguments \cite{Montero:2020icj}, indicating mechanisms beyond the anomaly \eqref{eq:anomaly} that break the 1-form center symmetry.
It would be interesting to find an explicit description for these breaking mechanisms.

\section{Discussion and Outlook}
\label{sec:discussion}

Using a mixed anomaly \eqref{eq:anomaly}, we have presented a necessary condition for an 8d ${\cal N}=1$ theory with given non-Abelian gauge algebras $\fkg_i$ at level $m_i$ to have a non-simply-connected gauge group $[\prod_i G_i]/Z$.
In combination with a set of Swampland criteria that restrict the gauge rank and the levels $m_i$, this condition rules out a vast set of possible gauge groups for 8d theories.
The constraints are especially powerful for theories of rank 18, where they reduce to known geometric properties of elliptic K3 surfaces.
As these properties control the global gauge group structure in F-theory compactifications, the anomaly provides a purely physical explanation for the intricate patterns of realizable gauge groups in F-theory.
The anomaly can further make predictions for inconsistent models in lower-rank cases, where the global gauge group structure in the corresponding string compactifications is yet to be explored systematically.

We stress that the absence of the anomaly \eqref{eq:anomaly} is only a necessary, but not sufficient condition for the gauge group to be $G/Z$.
Indeed, for F-theory constructions of the non-simply-connected gauge groups \eqref{eq:allowed_groups1}, there also exist K3 surfaces that realize the simply-connected versions in F-theory \cite{Shimada}.
There are also other instances where both $G$ and $G/Z$ are realized in different compactifications; this is also confirmed in the heterotic picture \cite{Font:2020rsk}.
As the center $Z$ in all these cases is non-anomalous, this is consistent with our findings.
At the same time, it is pointing toward additional breaking mechanisms, e.g., in terms of massive states charged under $Z$.
It would be interesting to investigate if these mechanisms are captured by an effective description involving the 1-form center symmetry.

There are also non-anomalous cases that have no realization in known classes of 8d string models.
A particular set of such cases are products $\frac{G_1}{Z_1} \times \frac{G_2}{Z_2}$ of anomaly-free factors, which would again be anomaly-free.
For example, the anomaly-free gauge group $[SU(5)^2/\bbZ_5] \times [SU(2)^4/\bbZ_2] = [SU(5)^2 \times SU(2)^4]/\bbZ_{10}$ as the non-Abelian part of a rank-18 theory (and, thus, all $m_i=1$) has no string realization.
As we have mentioned above, the F-theory geometry would forbid this case, because there is no $SU(n)$ factor with 10 dividing $n$.
Currently, we do not know an adequate physical argument providing the same restriction.
In terms of identifying gauged $\bbZ_\ell$ center symmetries, one plausible possibility is the existence of some mechanism that forces the presence of a $U(1)$ gauge factor into which, similarly to the hypercharge in the Standard Model, that $\bbZ_\ell$ embeds.
Such a theory would not be in contradiction to F-theory models, as center symmetries embedded in $U(1)$s have a different geometric origin \cite{Cvetic:2017epq} (see \cite{Cvetic:2015txa,*Cvetic:2018ryq,*Cvetic:2019gnh,*Cvetic:2020fkd} for direct implications for 4d particle physics models) not subject to the restriction \eqref{eq:anomaly_cancellation_rank18}.
Moreover, in 8d F-theory, there are additional sources for $U(1)$ factors (harmonic $(1,1)$-forms on K3's that are not algebraic), whose center-mixing with non-Abelian gauge factors needs further investigation.
To complete the geometric picture from the field-theoretic side, one must also extend the discussion of anomalies to include $U(1)$ gauge sectors, which we defer to future studies.

We further suspect that other discrete symmetries of the theory can interact non-trivially with 1-form center symmetries, leading to further constraints on the gauge group structure.
For example, it has been pointed out \cite{deBoer:2001wca} that the gauge symmetry of the $E_8 \times E_8$ heterotic string should be augmented by an outer automorphism $\bbZ_2$ exchanging the $E_8$ factors, so that the gauge group is $(E_8 \times E_8) \rtimes \bbZ_2$.
In fact, the 9d CHL string arises as the $S^1$-reduction with holonomies in this $\bbZ_2$.
Such an identification would also be possible for, e.g., $[SU(2)^4 / \bbZ_2] \times [SU(2)^4 / \bbZ_2]$, all at $m_i = 1$, which in 8d is free of the anomaly \eqref{eq:anomaly}, but not realized in terms of a string compactification.
If one could establish other field theory / Swampland arguments for why the $\bbZ_2$ outer automorphism must be gauged in this case, there could be other mixed anomalies involving the 1-form symmetries such that only a diagonal $\bbZ_2$ center survives, leading to the realizable $[SU(2)^8]/\bbZ_2$ theory.

Finally, to fully classify the global gauge group structure in 8d ${\cal N}=1$ theories based on the 1-form anomaly \eqref{eq:anomaly}, it will be important to have more stringent constraints on the possible levels $m_i$ for given simple gauge factors $G_i$.
While for rank-18 theories, \eqref{eq:anominfl} fixes all $m_i=1$, they cannot be fully determined by this method alone for rank-10 or -2 theories, and will require new tools and concepts to predict these independently from concrete string realizations.
Perhaps, new ideas can arise by establishing a connection between higher-form anomalies and the Swampland ideas \cite{Montero:2020icj} that also rule out certain non-simply-connected gauge groups.
These insights can hopefully lead to a complete understanding of the global gauge group structure, and prove full String Universality for non-simply-connected groups in 8d.

\begin{acknowledgments}
We thank Miguel Montero and Cumrun Vafa for helpful comments.
M.D.~and L.L.~further thank Fabio Apruzzi for valuable discussions and collaboration on related work \cite{Apruzzi:2020zot}.
M.D.~is grateful to Miguel Montero for valuable discussions.
We further thank the referee for pointing out the importance of the parameters $m_i$.
The work of M.C.~is supported in part by the DOE (HEP) Award DE-SC0013528, the Fay R.~and Eugene L.~Langberg Endowed Chair, the Slovenian Research Agency (ARRS No.~P1-0306), and the Simons Foundation Collaboration on ``Special Holonomy in Geometry, Analysis, and Physics,'' ID: 724069. The work of M.D.~is supported by the individual DFG grant DI 2527/1-1.
\end{acknowledgments}

\begin{appendix}

\section{Bounds on Coefficients}
\label{app:m}

In this appendix we briefly explain how anomaly inflow arguments can restrict the integer coefficients $m_i$ in \eqref{eq:anomcoup}. From \cite{Montero:2020icj} one knows that consistent theories in 8d have a gauge group with rank $\in \{18,10,2\}$. This means, that for a fixed rank the non-Abelian part of the gauge group captured by $G$ has to be supplemented by 
\begin{align}
n_{U(1)} = \text{rank} - \text{rank} (G) \,,
\end{align}
Abelian $U(1)$ factors. With the gauge algebra specified, one then has to restrict the prefactors $m_i$ such that the bound on the left-moving central charge \eqref{eq:anominfl} is satisfied.

\subsection{Rank 18}

For the case of a gauge group of rank 18, we note that the anomaly contribution is bounded from below by the rank of the corresponding group factor
\begin{align}
\frac{m_i \, \text{dim}(G_i)}{m_i + h_i} \geq \text{rank}(G_i) \,,
\end{align}
with the inequality satisfied for simply-laced groups at $m_i = 1$. This implies
\begin{align}
c_L \geq 18 \geq \sum \frac{m_i \, \text{dim}(G_i)}{m_i + h_i} + n_{U(1)} \geq \text{rank} = 18 \,,
\label{eq:ineqbound}
\end{align}
from which we see that for rank-18 cases the only gauge factors allowed are simply-laced $G_i$ with
\begin{align}
m_i = 1 \,,
\end{align}
i.e., the prefactors are fixed uniquely.

For reference, the values of $h$ and $\text{dim}(G)$ for allowed simple Lie groups $G$ in 8d are:
\begin{align}\label{tab:coxeter}
	\renewcommand{\arraystretch}{1.3}
	\begin{array}{|c||c|c|c|c|c|c|}
		\hline
		G & SU(n\geq 2) & Spin(n \geq 8) & Sp(n) & E_6 & E_7 & E_8 \\ \hline \hline 
		\text{dim} & n^2 - 1 & \tfrac{n(n-1)}{2} & n(2n+1) & 78 & 133 & 248\\ \hline
		h & n & n-2 & n+1 & 12 & 18 & 30 \\ \hline
	\end{array}
\end{align}

\subsection{Lower Ranks}

For lower ranks the possibilities for $(G_i, m_i)$ combinations increase since now the right-hand side of \eqref{eq:ineqbound} is smaller than 18. 
In particular, for smaller gauge groups or the case $\text{rank} = 2$ the combinatorics become more involved, which requires a case by case analysis beyond the scope of this work, that will be treated in more detail elsewhere.
Here, we make some simple observations about rank-10 cases that saturate the bound, which are also known to arise from the CHL string in 8d.

For the first example we consider the CHL string with non-Abelian gauge group $E_8 \times U(1) \times U(1)$.
This model has an $E_8$ current algebra at level $2$, and therefore
\begin{align}
\frac{m \, \text{dim}(G)}{m + h} + n_{U(1)}  = \tfrac{35}{2} \,,
\end{align}
which basically saturates the bound of 18, as there are no possible group/level contributing with $1/2$.
Higher levels of the current algebra with $m \geq 3$ are forbidden in this case.
Note that for the typical product subgroups with non-trivial quotient structure, such as $[E_7 \times SU(2)]/\bbZ_2$ or $[E_6 \times SU(3)]/\bbZ_3$, the anomaly \eqref{eq:anomaly} of the gauged center is indeed trivial.


In the second example we consider the CHL string with its maximal symplectic group, which is $Sp(10)$. The current algebra is at level 1, and one finds
\begin{align}
\frac{m \, \text{dim}(G)}{m + h} = \tfrac{35}{2} \,,
\end{align}
again saturating the allowed upper bound and prohibiting $m \geq 2$.
8d theories with level 1 groups in \eqref{eq:allowed_groups2} can potentially arise as subgroups, and will be investigated in detail in future work.

\section{Mordell--Weil Torsion and the Gauge Group in F-theory}
\label{sec:geometry}

8d ${\cal N}=1$ string compactifications with total gauge rank 18 can be described by F-theory compactified on elliptic K3 surfaces.
We refer to reviews \cite{Taylor:2011wt, Weigand:2018rez, Cvetic:2018bni} for a broader introduction, and focus in the following on two aspects key to discussion of global gauge group structure and center anomalies.
First, the non-Abelian gauge algebras $\fkg_i$ (associated to simply-connected groups $G_i$) are captured by reducible Kodaira-fibers of ADE-type $\fkg_i$ \cite{Vafa:1996xn,*Morrison:1996na,*Morrison:1996pp}.
Second, the global structure of the gauge group, $[\prod_i G_i]/Z$, is determined by the torsional part $Z$ of the Mordell--Weil group of sections \cite{Aspinwall:1998xj} (see especially \cite{Cvetic:2018bni} for a pedagogical introduction for this).

In general, the notation $G/Z$ requires a specification of the subgroup $Z \subset Z(G) = \prod_i Z(G_i)$ of the full center $Z(G)$.
In F-theory, this is determined by the intersection pattern between the generating sections of the Mordell--Weil group and the components of the $\fkg_i$-Kodaira fiber that form the affine Dynkin diagram of $\fkg_i$ \cite{Mayrhofer:2014opa,Cvetic:2017epq}.

For definiteness, we restrict ourselves to compactifications with $\fkg_i = \mathfrak{su}_{n_i}$, $i=1,...,s$, realized by K3 surfaces with only $I_{n_i}$ fibers.
Each reducible $I_{n_i}$ fiber consists of $n_i$ irreducible components that form a loop, reflecting the structure of the affine $\mathfrak{su}_{n_i}$ Dynkin diagram.
Starting with the affine node (determined by intersection with the zero-section) we label the components by $0,...,n_i-1$ as we go around the loop of the $i$-th fiber.
Then, an $\ell$-torsional section $\tau$ is \emph{uniquely} characterized by the tuple $(k_1,...,k_s)$ which labels the $k_i$-th component in the $i$-th fiber met by $\tau$ \cite{MirandaPersson}.
Moreover, one has $k_i \ell \equiv 0 \mod n_i$.

As explained in \cite{Mayrhofer:2014opa,Cvetic:2017epq}, $\tau = (k_1,...,k_s)$ corresponds precisely to the order $\ell$ element $(k_1,...,k_s) \in \prod_i \bbZ_{n_i} = Z(\prod_i SU(n_i))$.
This element acts trivially on all matter states of the F-theory compactification, hence giving rise to the gauge group $G/\langle \tau \rangle \cong G/\bbZ_\ell$.

The allowed combinations of $G$ and $\bbZ_\ell$ is heavily constrained geometrically by the following fact pertaining to intersection patterns between torsional sections and fiber components.
For a K3 $X$ with only $I_{n_i}$ fibers, the non-affine components of each fiber span a sublattice $R \subset H_2(X,\bbZ)$ with $\text{rank}(R) \leq 18$.
Then, one can associate to $R$ a so-called discriminant-form group \cite{MirandaPersson}
\begin{align}
	G_R \cong \bbZ_{n_1} \times ... \times \bbZ_{n_s} \, .
\end{align}
This group inherits from the lattice structure of $H_2(X,\bbZ)$ a quadratic form
\begin{align}\label{eq:quadratic_form}
	q: G_R \rightarrow \mathbb{Q}/\bbZ \, , \quad (x_1, ..., x_s) \mapsto \sum_{i=1}^s \frac{1-n_i}{2n_i} x_i^2 \mod \bbZ \, .
\end{align}
Notice that $G_R = \prod_{i=1}^s Z(SU(n_i)) = Z(G)$.
Then, by regarding a torsional section $\tau = (k_1,...,k_s)$ as an element of $G_R$, it can be shown \cite{MirandaPersson} that
\begin{align}\label{eq:isotropy_torsion_section}
	q(k_1,...,k_s) = \sum_{i=1}^s  \frac{1-n_i}{2n_i} k_i^2 \equiv 0 \mod \bbZ \,.
\end{align}
This precisely reproduces the constraint for the absence of the anomaly \eqref{eq:anomaly} involving the $\bbZ_\ell$ center 1-form symmetry with $m_i = 1$ for all $G_i$.

\end{appendix}

\bibliography{FM}

\end{document}